\documentclass[showpacs,preprintnumbers,amsmath,amssymb,APSl,prd,nofootinbib,superscriptaddress, twocolumn]{revtex4-1}

\usepackage{dcolumn}
\usepackage{bm}
\usepackage{ifpdf}
\usepackage{hyperref}
\usepackage{bm}
\usepackage{xcolor,color,graphicx,graphics}
\usepackage[spanish,english]{babel}
\usepackage[latin1]{inputenc}
\usepackage[OT1]{fontenc}
\usepackage{latexsym,amssymb,amsmath,amsfonts}
\usepackage{makeidx}
\usepackage{epsfig,subfigure}
\usepackage{natbib}
\usepackage{epstopdf}
\usepackage{mathrsfs}
\hypersetup{colorlinks=true, linkcolor=blue, citecolor=green}
\usepackage{enumerate}
\usepackage{xcolor}
 \usepackage{multirow}

\definecolor{red}{rgb}{1,0,0}

\def\+{^\dagger}

\def\<{\leftarrow}
\def\>{\rightarrow}

\def\({\left(}
\def\){\right)}

\newcommand{\bi}{\begin{itemize}} 				\newcommand{\ei}{\end{itemize}}
\newcommand{\benu}{\begin{enumerate}} 		\newcommand{\enu}{\end{enumerate}}
\newcommand{\bd}{\begin{dinglist}{0}}     \newcommand{\ed}{\end{dinglist}}
\newcommand{\bfig}{\begin{figure}[htbp]}  \newcommand{\efig}{\end{figure}}
        			
\newcommand{\bc}{\begin{center}} 				  \newcommand{\ec}{\end{center}}
\newcommand{\be}{\begin{equation}} 				\newcommand{\ee}{\end{equation}}
\newcommand{\bsub}{\begin{subequations}}  \newcommand{\esub}{\end{subequations}}
\newcommand{\ba}[1]{\begin{array}{#1}} 		\newcommand{\ea}{\end{array}}

\begin{document}
\title{New light rings from multiple critical curves as observational signatures of black hole mimickers}

	\author{Gonzalo J. Olmo} \email{gonzalo.olmo@uv.es}
\affiliation{Departamento de F\'{i}sica Te\'{o}rica and IFIC, Centro Mixto Universidad de Valencia - CSIC.
Universidad de Valencia, Burjassot-46100, Valencia, Spain}
\affiliation{Departamento de F\'isica, Universidade Federal da
Para\'\i ba, 58051-900 Jo\~ao Pessoa, Para\'\i ba, Brazil}
\author{Diego Rubiera-Garcia} \email{drubiera@ucm.es}
\affiliation{Departamento de F\'isica Te\'orica and IPARCOS,
	Universidad Complutense de Madrid, E-28040 Madrid, Spain}
	\author{Diego S\'aez-Chill\'on G\'omez} \email{diego.saez@uva.es}
\affiliation{Departamento de F\'isica Te\'orica, At\'omica y \'Optica, Campus Miguel Delibes, \\
Universidad de Valladolid UVa, Paseo Bel\'en, 7, 47011 - Valladolid, Spain}

\date{\today}
\begin{abstract}
We argue that the appearance of additional light rings in a shadow observation - beyond the infinite sequence of exponentially demagnified self-similar rings  foreseen in the Kerr solution - would make a compelling case for the existence of black hole mimickers having multiple critical curves. We support this claim by discussing three different scenarios of spherically symmetric wormhole geometries having two such critical curves, and explicitly work out the optical appearance of one such object when surrounded by an optically and geometrically thin accretion disk.

\end{abstract}

\maketitle

\section{Introduction}

For decades after Schwarzschild found its celebrated solution, the relativistic community had to deal with the conceptual novelties that the presence of (what is nowadays known as)  an event horizon raised. It was not until the Kerr(-Newman) family of axisymmetric electrovacuum solutions of Einstein's field equations of General Relativity (GR) was found \cite{Waldbook}, together with the understanding of gravitational collapse \cite{Joshibook}, that black holes (BHs) were accepted as a reliable representation of the final fate of massive enough fuel-exhausted stars. The Kerr solution has passed numerous tests for its astrophysical viability, most notoriously those based on electromagnetic radiation \cite{Bambi:2015kza}, which recently include the shadows surrounding the object \cite{EventHorizonTelescope:2019dse}. The latter corresponds to the dark silhouette of a BH when illuminated by an accretion disk \cite{Luminet:1979nyg,Narayan:2019imo}, where the gravitational deflection of light rays approaching an unstable circular orbit (a critical curve) defines the boundary of the central brightness depression. Moreover, by assuming the disk to be optically thin, this boundary is made up of a set of strongly lensed and gravitationally redshifted light rings superimposed on the direct emission \cite{Falcke:1999pj}. 
 
The current constraints on shadows shapes set by the EHT Collaboration measurements \cite{EventHorizonTelescope:2019dse} point to their compatibility with the rough expectations of the Kerr solution. However, such shadows might be produced by other compact objects having a critical curve \cite{Guerrero:2021ues}.  Indeed, since Nature could breed creatures beyond our imagination, it is pertinent to characterize the zoo of {\it BH mimickers} \cite{Johnson-Mcdaniel:2018cdu} able to cast a shadow. Moreover, since two different BH (mimickers) of comparatively similar masses are able to cast  the same shadow \cite{Lima:2021las}, there is a major effort in conceiving new physical signatures capable to distinguish unmistakably a BH from a mimicker \cite{Cardoso:2019rvt,Guerrero:2021ues}. The investigation of such signatures is of particular interest given the present great opportunity to carry out precise tests of the Kerr solution in the strong field regime \cite{EventHorizonTelescope:2020qrl}. The main aim of the present Letter is to bring forward one of such signatures, which is driven by the existence of multiple critical curves in some BH mimickers. \\
\indent For the sake of this work, we shall consider one of the leading proposals for BH mimickers, namely, the wormhole (WH) paradigm. Likewise BHs in the past, WHs have awaken a special interest by decades in the community. Usually interpreted as structures allowing to connect far away regions, rather than a GR-framed solution they are a generic possibility in gravitational theories based on a curved space-time. Since they naturally restore geodesic completeness \cite{Carballo-Rubio:2019fnb}, which is a deep-rooted problem in GR BHs,  and may even shed new light on quantum information problems \cite{Maldacena:2013xja}, they are worth bearing in mind. Hence, we shall discuss three different WH scenarios where two critical curves are present. We shall argue that the presence of the new critical curve in optically thin disks generically brings in additional light rings not present in the Kerr solution, some of which may yield a non-negligible contribution to the total luminosity of the object. This claim will be further strengthened by explicitly working out the shape and luminosities of such new light rings in a toy-model consisting of a WH extension of the Schwarzschild solution, illuminated by three different emission profiles of a geometrically thin accretion disk. The observation of such new light rings would act as compelling evidence for the existence of this kind of BH mimickers.

\section{Critical curves, ray-tracing and thin accretion disk}

We consider a spherically symmetric (WH) geometry written in suitable coordinates as:
\be
 ds^2=-A(x)dt^2+A^{-1}(x)dx^2+r^2(x)d\Omega^2\ ,
 \label{MetricWH}
 \ee
 where the (non-vanishing) area of the two-spheres $S=4\pi r^2(x)$ is parametrized in terms of the coordinate $x \in (-\infty,+\infty)$. Though broadly astrophysical compact objects are expected to rotate, the size and shape of the shadow and its boundary are weakly dependent on the spin and the orientation of the observer \cite{Johannsen:2010ru,Perlick:2021aok,Gralla:2019xty}, so the spherical symmetry restriction will allow us a cleaner scenario to capture the main observational novelties brought by the existence of multiple critical curves. For a null geodesic and taking $\theta=\pi/2$, a freedom granted by the spherical symmetry restriction, the corresponding geodesic equation  reads
\begin{equation} \label{eq:geo1}
\dot{x}^2=\frac{1}{b^2}-V_{eff}(x) \ ,
\end{equation}
where $\dot{x} \equiv dx/d\lambda$ with $\lambda$ being an affine parameter and  $b=L/E$  the impact parameter ($E=A\dot{t}$ and $L=r^2(x)\dot{\phi}$ being the photon's conserved energy and angular momentum, respectively) whereas the effective potential is given by:
\be
V_{eff}=\frac{A(x)}{r^2} \ .
\label{EffePotent}
\ee
Unstable circular orbits (the so-called photon sphere) will exist as far as $\dot{x}=0$ for every $\lambda$, such that the right-hand side of equation (\ref{eq:geo1}) vanishes at the circular radius, given by  $x_m$, and representing a maximum of the effective potential (\ref{EffePotent}), i.e., $V_{eff}(x_{ps})=1/b^2, V_{eff}'(x_{ps})=0,V_{eff}''(x_{ps})<0$. This defines  a critical impact parameter
\begin{equation}
b_c=\sqrt{\tfrac{r^2(x_{ps})}{A(x_{ps})}} \ ,
\end{equation}
where a light ray would turn the object an infinite number of times. For the Schwarzschild BH, one has $b_c=3\sqrt{3}M$ with $r_{ps}=x_{ps}=3M$. 

We consider an accretion disk whose emission is confined to the equatorial plane. In the optically thin limit, which seems to be a  reliable enough description for the accretion disk of supermassive BHs \cite{Johnson:2015iwg}, each pixel on the observer's screen collects the integrated emission along a given null geodesic. Thus, a light ray may intersect the disk several times in their winding around the object, picking up additional brightness according to the emission profile of the disk. Astrophysical observations of the intensity on the observer's screen are thus computed by backward ray-tracing the null geodesics equation (\ref{eq:geo1}), suitably rewritten as the variation of the azimuthal angle with respect to the radial coordinate:
\begin{equation} \label{eq:geo2}
\frac{d\phi}{dx}=\frac{b}{r^2(x)\sqrt{1-\frac{b^2 A(x)}{r^2(x)}}} \ .
\end{equation}
This procedure splits the light rays issued from the observer into two classes: those with $b>b_c$ are deflected at a minimum distance to asymptotic infinity, and those with $b<b_c$ spiral down until meeting the event horizon (if any). The image on the observer's screen is thus formed by an infinite set of self-similar light rings characterized by the number of photon (half-)orbits in their winding around the compact object. Since such rings will be exponentially thinner and dimmer \cite{Luminet:1979nyg}, the image of the object will be largely dominated by the direct emission (a single intersection with the disk) appearing as a thick bright lump of radiation. As discussed in \cite{Gralla:2019xty}, only the lensing ring (two intersections with the front and back of the disk, respectively) produces a neatly visible additional bright ring, while for the sum of higher-orbit contributions (trajectories intersecting the disk at least thrice) only the first sub-ring is barely visible (what the authors of \cite{Gralla:2019xty} dub as ``photon ring", dismissing all the subsequent orbits).  It is worth pointing out that, in those models of accretion disks whose inner edge extends below the critical curve (up to the horizon), there will be emissions from inside that region towards the observer's screen, effectively enlarging the spaces of direct/lensed/photon ring trajectories inside the region $b<b_c$. Therefore, the optical appearance of the object via its pattern of luminous and dark regions will be highly influenced by the emission profile of the disk. At the same time, and despite their relative dimness, the fact that  the location of  higher-order rings are very sensitive to the background geometry rather than to the details of the accretion disk, makes them suitable signatures to probe the metric itself \cite{Wielgus:2021peu}.

The geometrically thin accretion disk is described by the radiative transfer (Boltzmann) equation for unpolarized photons \cite{Gold:2020iql}. Neglecting scattering and absorption effects, and assuming a source with an isotropic emission in the rest frame of the disk with specific intensity $I^{em}_{\nu}=I(x)$ and a frequency-independent emissivity,  one finds that  $I_{\nu}^{em}/\nu^3$ is conserved along a null geodesic \cite{Eichhorn:2021iwq}.  In the observer's frame with photon's frequency $\nu'$, its intensity is given by $I^{ob}_{\nu'}=(\nu'/\nu)^3I^{em}_{\nu}$. For the geometries considered here this implies $I^{ob}_{\nu'}=A(x)^{3/2}I^{em}_{\nu}$, so that the integrated intensity becomes $I^{ob}= \int  d\nu' I_{\nu'}^{ob}=A^2(x)I(x)$. The total intensity comes from taking into account all the possible intersections of every photon with the disk as:
\begin{equation}
I=\sum_{m=1}^3 A^2(x)I(x)_{\vert_{x=x_{m}(b)}}\ ,
\end{equation}
where the so-called transfer function $x_m(b)$ (with $m=1,2,3$ for direct/lensed/photon ring trajectories) collects the relation between the emission radius of the disk and the impact parameter $b$, with its slope encoding the corresponding demagnification of the image. 

Going beyond Schwarzschild spacetimes, more than one critical curve might naturally arise. In such a case light rays will be able to spiral around the additional critical curve(s) following a similar description of direct/lensed/photon ring trajectories. Moreover, in the intermediate region between every pair of critical curves the (local) minimum of the potential will act as an anti-photon sphere, around which light rays may oscillate in elliptical orbits \cite{Cardoso:2014sna}. Therefore, a complex pattern of new trajectories will arise in such objects with multiple critical curves, in such a way that, depending on the emission profile of the disk and, in particular, on how deeply it penetrates inside the new critical curve(s), one would expect the appearance of new luminous and dark regions associated to the dominant sources of emission (direct/lensed/photon ring). In the next section we shall present three scenarios of WHs having two critical curves and work out the new pattern of light rings for one of such objects using three toy-models of accretion disk emissions.

\section{Light rings and shadows from WHs with multiple critical curves}

\subsection{ Absolute maximum of the potential at the center}

\begin{figure}[t!]
\includegraphics[width=8.0cm,height=4.7cm]{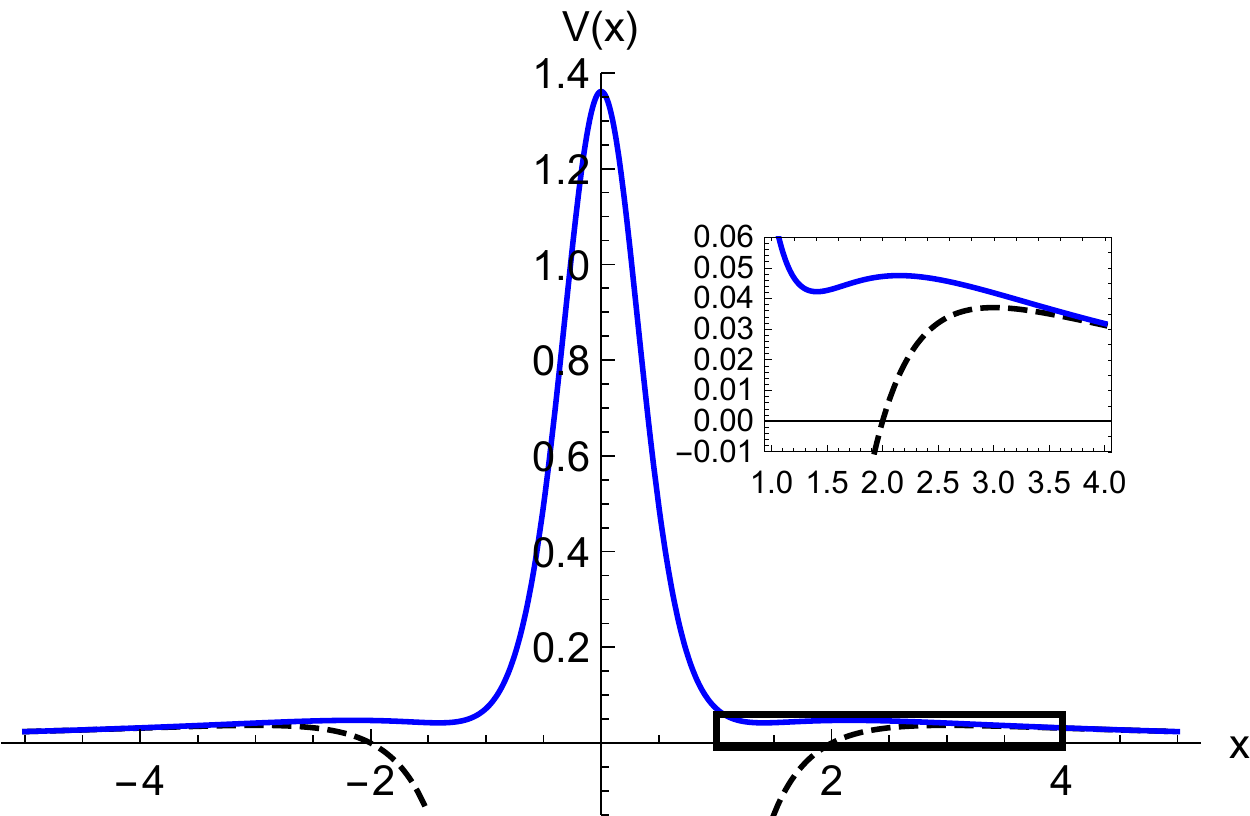}
\caption{The effective potential $V_{eff}(x)=A/r^2$ for the model (\ref{eq:metric}) with $a/M=6/7$ (blue). The inset panel depicts  the outermost (local) maximum, while the absolute maximum is located at the throat $x=0$, a feature which is missing in the Schwarzschild solution (dashed black).}
\label{fig:pot}
\end{figure}

Let us consider the uniparametric toy-model defined by 
\begin{equation} \label{eq:metric}
A(x)=1-\frac{2Mx^2}{(x^2+a^2)^{3/2}} \hspace{0.2cm} ; \hspace{0.2cm} r^2(x)=x^2+a^2
\end{equation}
which recovers the Schwarzschild solution in the limit $x \to \infty$, so $M$ is the mass of the object as seen by a far away observer. This toy-model is actually an extension of the well known Bardeen's solution \cite{Bardeen} where the radial function is infused with a bouncing behaviour at $x \to 0$ thanks to the presence of the parameter $a$, which in turn restores geodesic completeness and removes the presence of curvature divergences everywhere. We shall take a theory-agnostic approach, in which (\ref{eq:metric}) may arise in GR (supported e.g. by some non-linear theory of electrodynamics) or in some modified gravity theory \cite{Lobo:2020ffi}. This family of solutions describes either two-horizon BHs with a hidden WH inside it ($0<a/M \leq 4\sqrt{3}/9$) or traversable WHs otherwise. The corresponding effective potential has always an absolute maximum at $x=0$ (the throat), while a second (local) maximum is present for $ a/M \leq 2\sqrt{5}/5$. Hence, within $4\sqrt{3}/9<a/M \leq 2\sqrt{5}/5$ there are two critical curves \cite{Tsukamoto:2021caq}. For the sake of this work we shall choose $a/M=6/7$, whose corresponding potential is depicted in Fig. \ref{fig:pot}. The full set of images for every possible configuration within this family will be treated in a separate publication.

\begin{figure}[t!]
\includegraphics[width=8.0cm,height=4.7cm]{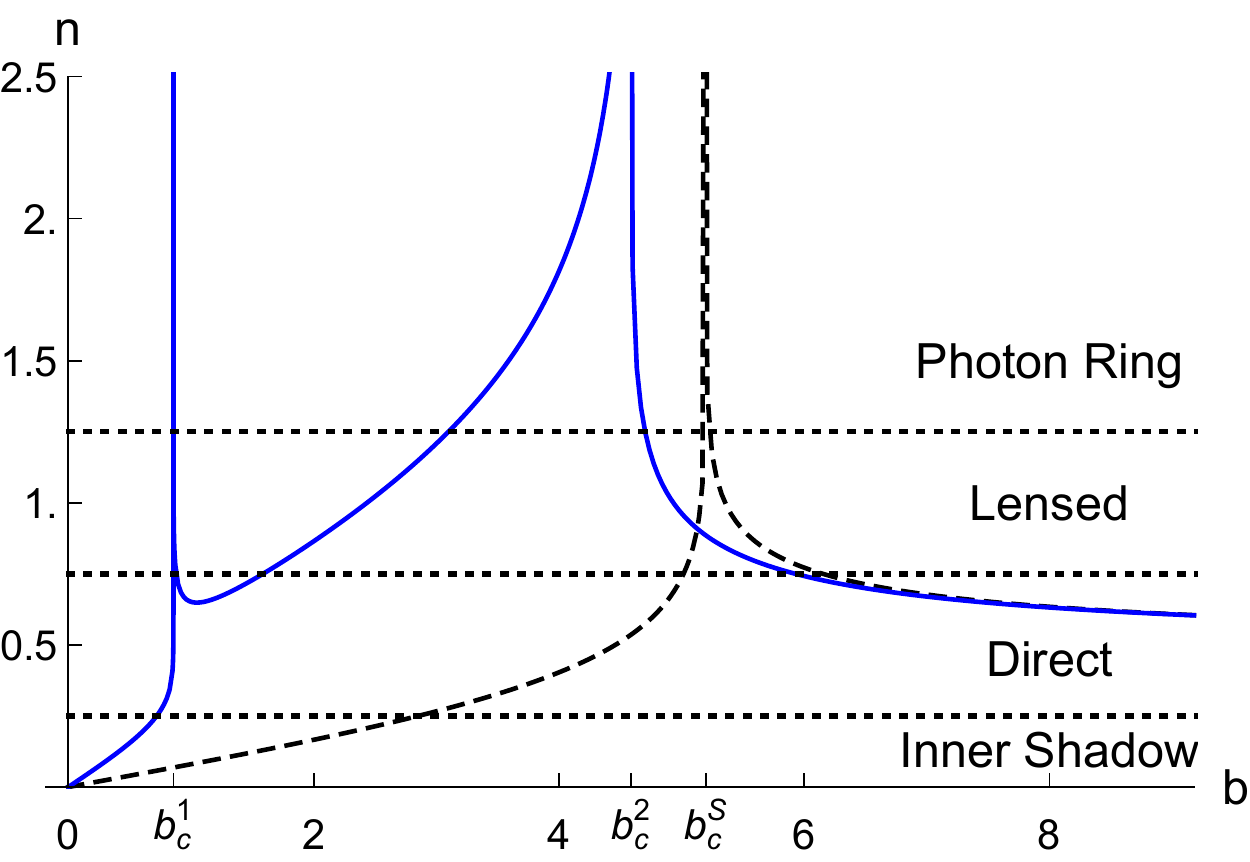}
\caption{The number of half-orbits (in blue) as a function of $b$ (here $M=1$), as compared to the Schwarzschild solution (dashed black). The two critical curves $b_c^1,b_c^2$ drive the direct/lensed/photon ring emissions, whose corresponding regions are enclosed by the horizontal dashed lines. We also depict the inner shadow limit, $b_{is}<b_c^1$, to be discussed later.}
\label{fig:orbits}
\end{figure}

\begin{figure}[t!]
\includegraphics[width=0.5\textwidth]{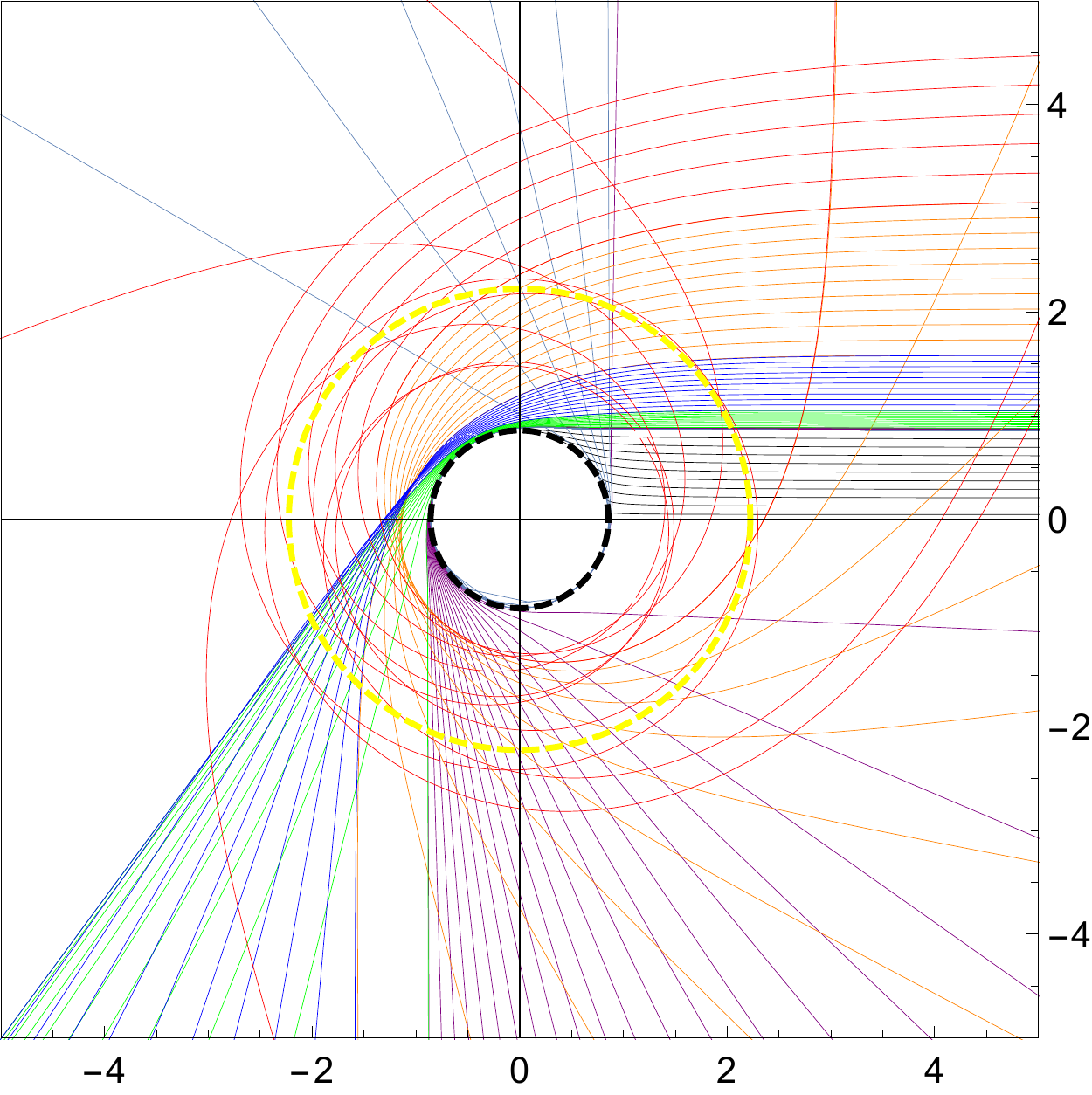}
\caption{Ray-tracing of the trajectories ($M=1$) within the two critical curves, $b_c^1<b<b_c^2$, where the different colours are identified with their photon ring/lensed/direct character according to Fig. \ref{fig:orbits}. The dashed yellow and black circumferences correspond to the outer and inner critical curves, while the black trajectories denote the inner shadow limit, $b_{is}<b_c^1$.}
\label{fig:raytracWH}
\end{figure}

\begin{figure}[t!]
\includegraphics[width=0.5\textwidth]{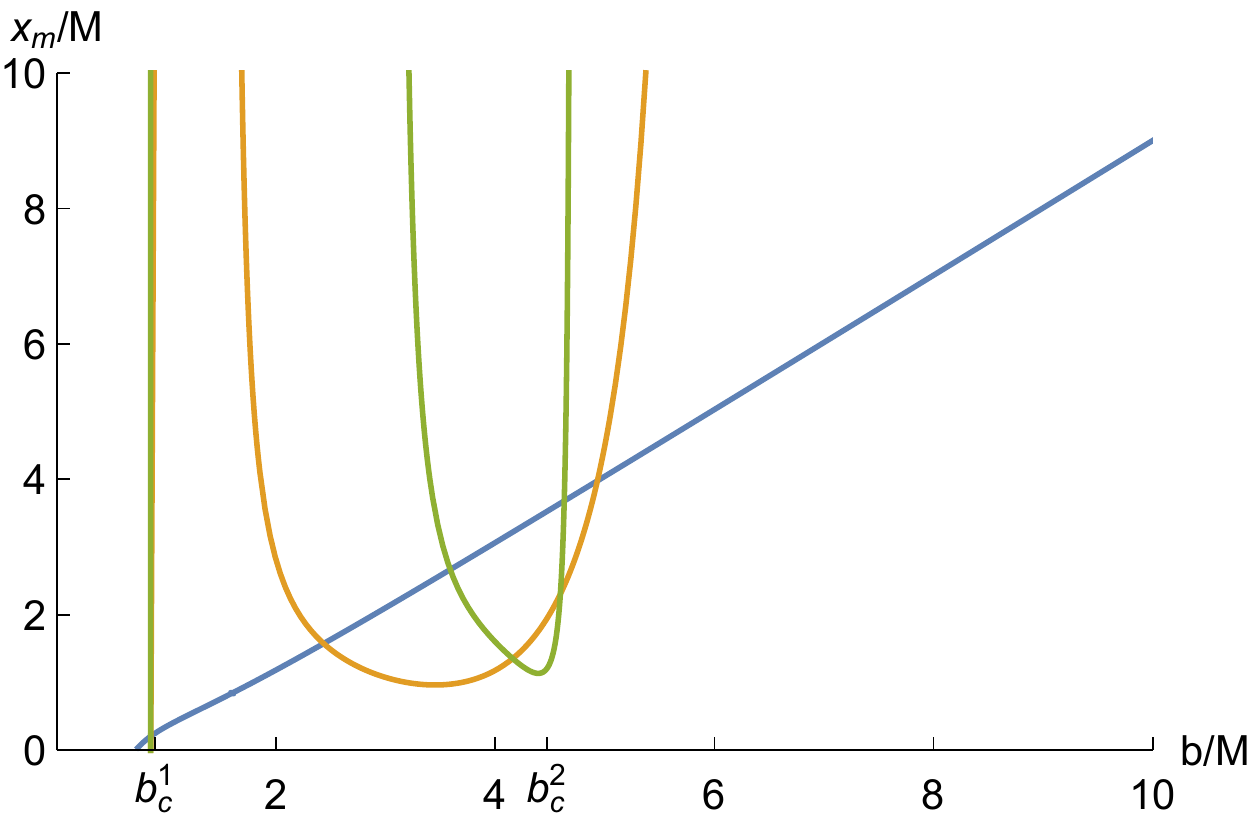}
\caption{The transfer function $x_m(b)$ for the direct (blue), lensed (orange) and photon ring (green) emissions. }
\label{fig:transfuncwh}
\end{figure}

We implement the ray-tracing procedure by considering a disk located on the equatorial plane (so that the observer lies on the north pole). The (normalized) number of (half-)orbits is defined by $n \equiv \phi/2\pi$, so that light rays in straight motion (outside the outer critical curve) is defined by $n=1/2$, while direct/lensed/photon ring trajectories correspond to $1/2<n<3/4$, $3/4<n<5/4$, and $n>5/4$, respectively.  Since we have now two critical curves with corresponding impact parameters $b_c^1/M=6/7$ and $b_c^2/M \approx 4.589$, these three types of trajectories arise around each of them, whose number of half-orbits is depicted in Fig. \ref{fig:orbits}. The two critical curves appear in this plot as two spikes such that in the region $b_c^1<b<b_c^2$ the description of the direct/lensed/photon ring emissions strongly deviates from the Schwarzschild spacetime expectations, as shown in Fig. \ref{fig:raytracWH}. The expected contribution of such emissions to the image within the different impact parameter regions is hinted by the transfer function $x_m(b)$ depicted in Fig. \ref{fig:transfuncwh}.

We consider next three accretion disk canonical toy-models where the maximum of the emission is attained at some relevant surface: i) innermost stable circular orbit for time-like observers, ii) (outer) critical curve, and iii) WH's throat, respectively, being zero in the interior region to them, while smoothly falling off outwards (see \cite{Guerrero:2021ues} for details of these models). This is intended to simulate different types of (geometrically thin) disks or, alternatively, different stages in the temporal evolution of the same disk, and to collect and integrate the luminosity associated to the direct/lensed/photon ring trajectories of the two critical curves in different ways. 

\begin{figure*}[t!]
\includegraphics[width=5.9cm,height=5.5cm]{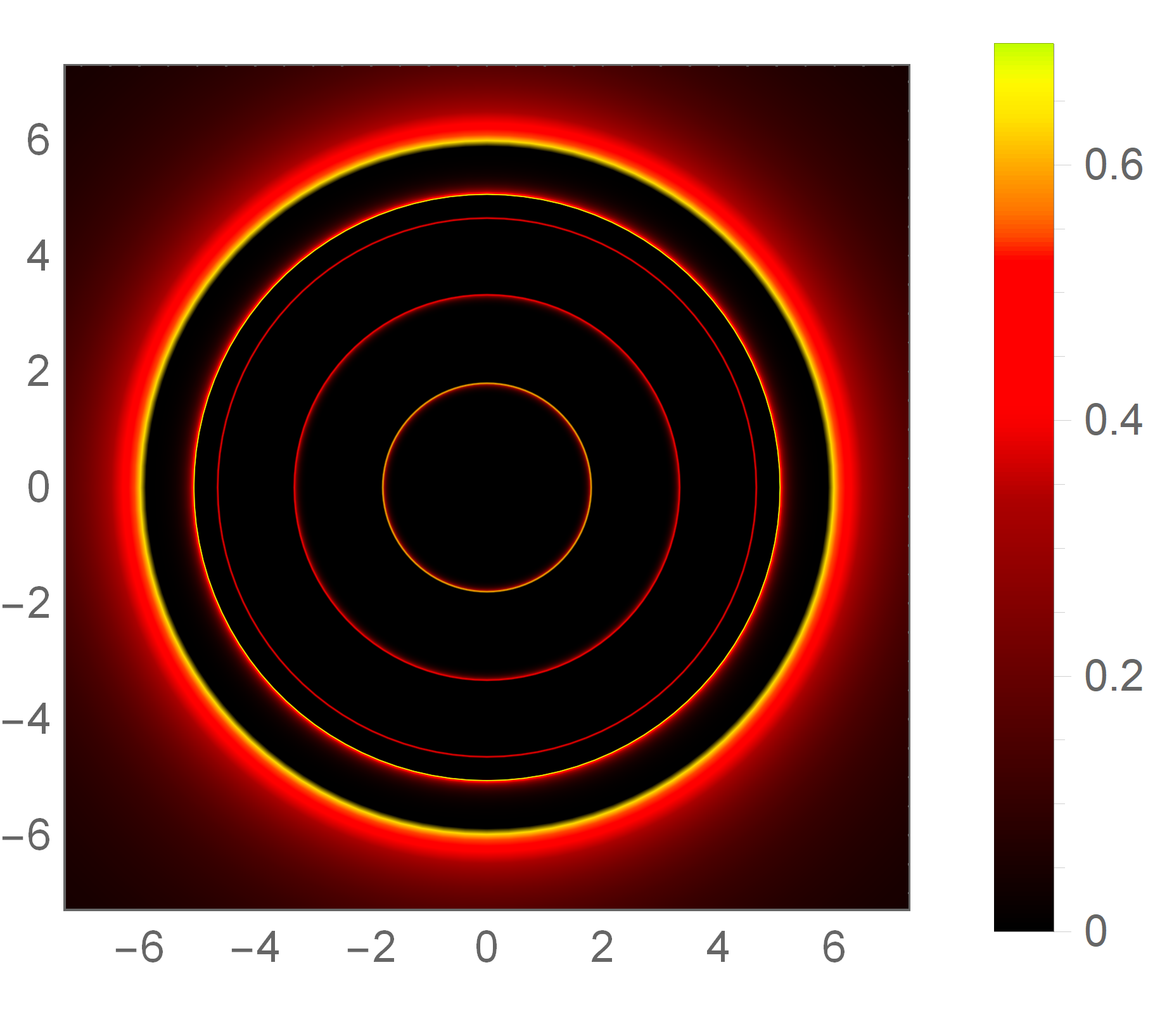}
\includegraphics[width=5.9cm,height=5.5cm]{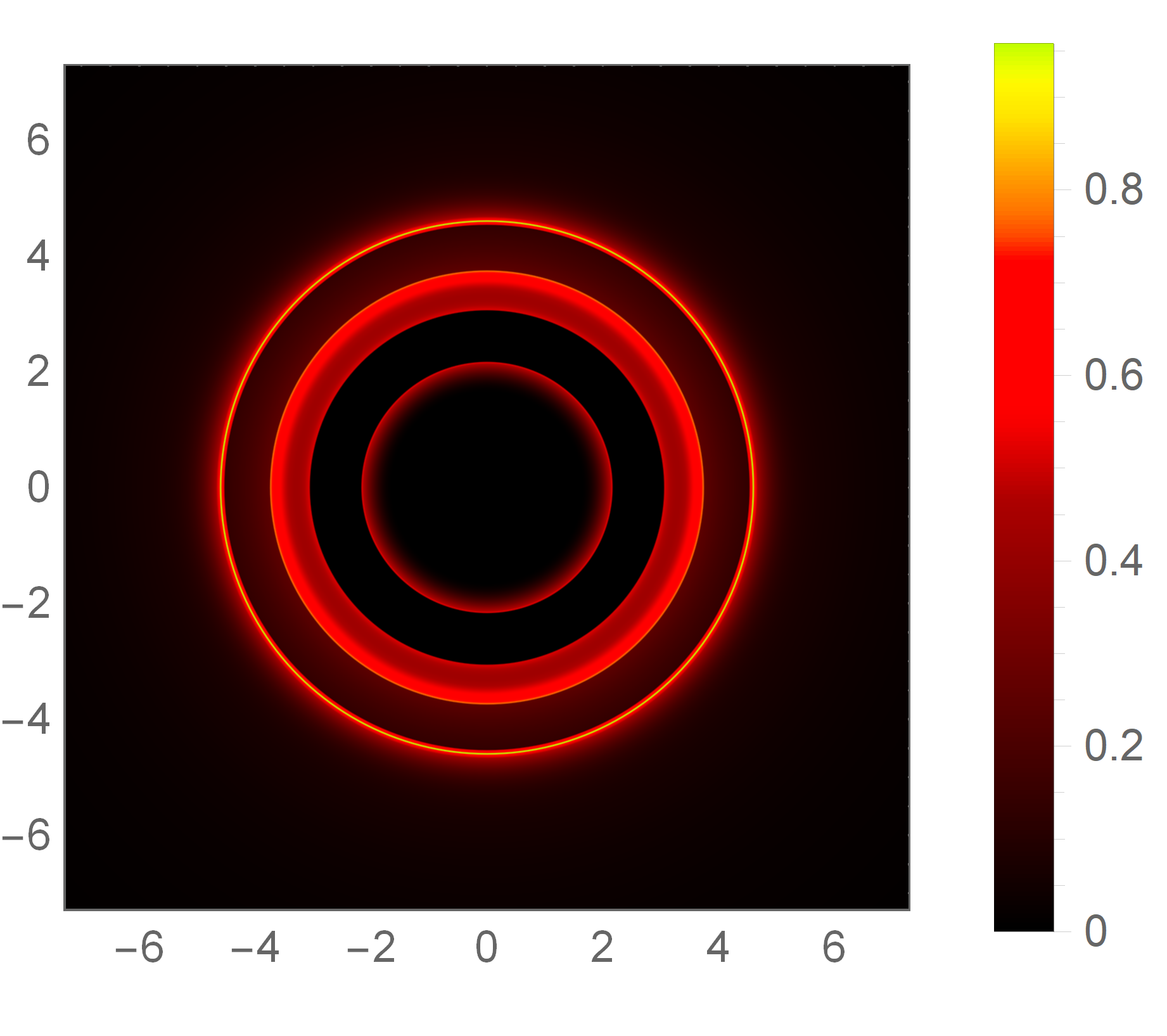}
\includegraphics[width=5.9cm,height=5.5cm]{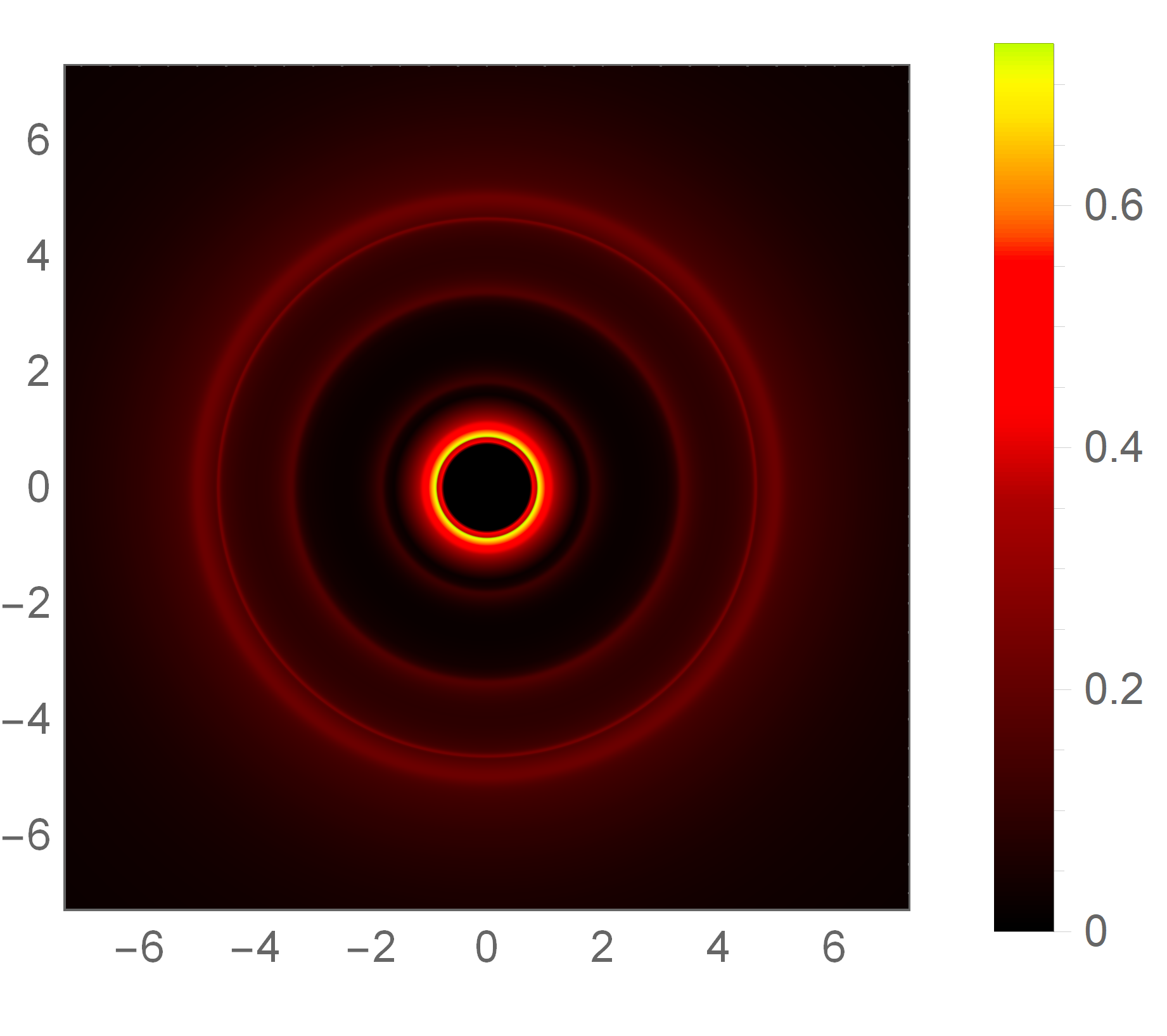}
\caption{The optical appearance (viewed in face-on orientation) of the traversable WH (\ref{eq:metric}) with $a=6/7$ and $M=1$ for an accretion disk whose inner edge extends to the innermost stable circular orbit for time-like observers (left), to the outer critical curve (middle), and all the way down to the WH throat (right). Three extra rings (one not visible) appear inner to those present in the Schwarzschild solution, associated to the new lensed/photon ring emissions. Depending on the accretion disk model, these rings can be neatly separated from each other (left) or superimposed with the direct emission (middle and right).}
\label{fig:shadowWH}
\end{figure*}

The optical appearances for these three accretion disk models are depicted in Fig. \ref{fig:shadowWH}. In the first model (left panel) we see five concentric neat circular rings. The three outermost ones are akin to the standard light rings of the Schwarzschild solution (associated to the direct/lensed/photon ring trajectories of the outer critical curve), slightly modified in their corresponding impact parameters and luminosities due to the geometry (\ref{eq:metric}). The additional two visible light rings are associated to photon ring/lensed trajectories of the inner critical curve, collecting $\sim 2.8\%$ and $\sim 2.5 \%$ of the total luminosity, respectively. Moreover, there is yet another (innermost) light ring associated to further photon ring emission of the inner critical curve but it is so dimmed ($\sim 10^{-6}$) that it is not visible at naked eye in this figure. In the second model (middle panel) the disk emission induces an overlapping between several contributions of these trajectories, abruptly changing the distribution of bright and dark regions. Now the innermost visible contribution is a new lensing ring associated to the inner critical curve accounting for a $\sim 9.2\%$ of the luminosity budget, the latter being largely dominated by the extended bright central ring (a mixing of direct and inner photon ring emissions), with another visible thin ring in the outermost edge. Finally, in the third model (right panel) there are several blurred rings with intermediate dark-brown regions and the image is dominated by the concentrated bright ring of radiation, where different contributions associated to both critical curves are combined, enclosing a central black region greatly diminished in size.

The bottom line of the discussion above is that, while a given background geometry is the solely actor determining the location of the critical curve(s), very different images are produced depending on the assumed emission profile of the accretion disk, inducing different degrees of superposition of the individual light rings on its interaction with the geometry of the critical curve(s). Let us also point out here that the model (\ref{eq:metric}) and its parameter are chosen on purpose in order to grossly exaggerate the features brought by the presence of the new critical curve as compared with the expected images of the GR (Schwarzschild) BH, though much less radical proposals do exist in the market \cite{Gan:2021pwu}.

\subsection{Accretion disk on the opposite side of the throat} 

For a BH, when the inner edge of the disk extends up to the horizon, the size of the central brightness depression is reduced towards the {\it inner shadow}, defined as the impact parameter $b_{is}<b_c^1$ below which a light ray terminates on the event horizon without crossing the equatorial plane once ($n<1/4$ in Fig. \ref{fig:orbits}). The fact that no light ray can be originated neither from inside the horizon nor from the accretion disk itself makes this region become an utter brightness depression. Moreover, being a feature uniquely ascribed to the background geometry, simultaneous observations of it and of the critical curve(s) might help to break the degeneracy of the shadow's caster parameters  \cite{Chael:2021rjo}. 

Let us now consider a traversable WH described by a symmetric potential with respect to the throat, with a single critical curve on each side. This can be achieved, for instance, in thin-shell scenarios by gluing together two patches of space-time $\mathcal{M}^{\pm}$ at the shell (the throat), $\Sigma =\mathcal{M}^+ \cap \mathcal{M}^-$ via a suitable junction conditions formalism in different theories of gravity \cite{Senovilla:2013vra,Olmo:2020fri}. Let us place our observer on $\mathcal{M}^+$, and assume an accretion disk on $\mathcal{M}^-$ only. Light rays coming from the $\mathcal{M}^-$ side with impact parameter $b<b_{is}$ will reach the mouth without having any intersection with the disk, and cross it to reach the observer's screen on $\mathcal{M}^+$. Therefore, unless the back-reaction of the matter fields threading the shell $\Sigma$ is capable to significantly alter the light ray's impact parameter, the image in the region $b<b_{is}$, despite not being pitch black, will only be given by those rays coming from the $\mathcal{M}^-$ sky, not capable to yield additional light rings. However, if there is an accretion disk on $\mathcal{M}^-$ whose emission extends inside the critical curve there, some of those light rays with $b_{is}<b<b_{c}$ may intersect that disk several times before crossing the throat to yield additional light rings on the $\mathcal{M}^+$ observer's screen within that impact parameter region. Moreover, if there is another accretion disk on $\mathcal{M}^+$ whose emission also extends inside the critical curve on that side, one would expect a superposition of the images originated on both sides on top of the direct emission(s). Likewise in the example above of the absolute maximum of the potential at the throat, whether these images will resolve in neatly separated light rings will be model-dependent both on the background geometry as on the features of the accretion disks on $\mathcal{M}^{\pm}$, as shown, for instance, in the two rotating WH geometries considered in \cite{Paul:2019trt}. 

\subsection{Reflection-asymmetric critical curves} 

Let us assume now the effective potential to have different shapes on $\mathcal{M}^{\pm}$. These {\it reflection-asymmetric WHs} have been considered recently in the literature \cite{Wielgus:2020uqz,Guerrero:2021pxt,Peng:2021osd}. Suppose again the observer to be located on $\mathcal{M}^+$ and the asymmetric potential to have its absolute maximum on $\mathcal{M}^-$. Now there is a new set of light trajectories within the range of impact parameters $b_c^{-}<b<b_c^+$, with $b_c^{\pm}$ the critical impact parameters associated to the maxima of the potentials on $\mathcal{M}^{\pm}$, respectively. Such trajectories  correspond to light rays originated on $\mathcal{M}^+$, which cross the throat $\Sigma$, hit the potential slope on $\mathcal{M}^-$, and get reflected back to $\mathcal{M}^+$. Therefore, whenever the emission of the accretion disk penetrates this region (on $\mathcal{M}^-$), additional light rings will appear associated to lensed/photon ring trajectories, again depending on the emission profile of the disk.

\section{Conclusion and remarks}

Though the application of Ockham's razor bids us to admit that every time we observe a shadow, a (Kerr) BH is the most likely candidate, the difficulty to disentangle the effects of the background geometry and the accretion disk modelling on such an image  \cite{Vincent:2020dij} drives us to consider alternative scenarios for the shadow caster, where clear cut observational signatures may arise. In this regard, the next generation of experiments together with new analytical tools might be capable to distinguish among those effects \cite{Lara:2021zth}. In this Letter we have argued that within the large zoo of BH mimickers, those having multiple critical curves (illustrated here with certain WH geometries)  in optically thin scenarios generically display additional light rings driven by the new lensed/photon ring emissions associated to the new critical curve(s). Using geometrically thin toy-models of the accretion flow we have shown that such new rings can be located in far-removed regions of the impact parameter space, and contribute non-negligibly to the total luminosity, even significantly larger than those of the infinite sequence of strongly-lensed rings associated to the standard critical curve of the Schwarzschild (Kerr) solution.  While these features could facilitate their detection, depending on the emission properties of the  disk these new light rings can be superimposed with the direct emission as well as with the canonical rings. This can spoil a neat separation of the (old and new) light rings, forming instead a reduced number of rings with wider amplitudes and larger combined luminosities. As a consequence, the optical appearance of any such mimickers may strongly deviate from the Schwarzschild BH one (as illustrated with the images presented in Fig. \ref{fig:shadowWH}).  Since very long baseline interferometry could be able to resolve out the diffuse but sharp contributions from higher-order light rings in the near future \cite{Johnson:2019ljv}, the observation of any such new rings from multiple critical curves could make a compelling case for the existence of non-Kerr beasts in the cosmic zoo.

\section*{Acknowledgements} 

 DRG is funded by the \emph{Atracci\'on de Talento Investigador} programme of the Comunidad de Madrid (Spain) No. 2018-T1/TIC-10431,  and acknowledges further support from the FCT projects No. PTDC/FIS-PAR/31938/2017 and PTDC/FIS-OUT/29048/2017. DS-CG is funded by the University of Valladolid (Spain), Ref. POSTDOC UVA20. This work is supported by the Ministerio de Ciencia e Innovaci\'on (Spain) projects Nos.  FIS2017-84440-C2-1-P, PID2019-108485GB-I00, PID2020-116567GB-C21, and  PID2020-117301GA-I00, the project PROMETEO/2020/079 (Generalitat Valenciana), and the Edital 006/2018 PRONEX (FAPESQPB/CNPQ, Brazil, Grant 0015/2019). This article is based upon work from COST Action CA18108, supported by COST (European Cooperation in Science and Technology).

\end{document}